\documentstyle[epsfig]{ta}
\begin{document}

\title{%
Search for continuous gravitational wave signals
from sources in binary systems}

\author{Sanjeev V.\ DHURANDHAR \\
{\it Max Planck Institut f\"{u}r Gravitationsphysik\\
Albert Einstein Institut \\
Am M\"{u}hlenberg 1, Golm \\
D-14476, Germany, sanjeev@aei-potsdam.mpg.de}\\
Alberto VECCHIO\\
{\it Max Planck Institut f\"{u}r Gravitationsphysik\\
Albert Einstein Institut \\
Am M\"{u}hlenberg 1, Golm \\
D-14476, Germany, vecchio@aei-potsdam.mpg.de}}

\maketitle

\section*{Abstract}

We analyze the computational costs of searches 
for continuous monochromatic gravitational waves 
emitted by rotating neutron stars orbiting a companion object. 
As a function of the relevant orbital parameters, we address 
the computational load involved in targeted 
searches, where the position of the source is known;
the results are applied to 
known binary radio pulsars and Sco-X1. 

\section{Introduction}

The search for continuous wave (CW) sources --
rapidly rotating neutron stars (NS) that emit quasi-monochromatic gravitational
waves (GW's) -- is one of the most computationally intensive tasks 
for the data analysis of GW detectors. Surveys of wide areas of the
sky and/or large frequency and spin-down ranges are computationally bound (Brady et al., 1998);
the only viable strategy is to set 
up hierarchical algorithms, where "coherent" and "incoherent" 
stages are alternated in order to maximize
the signal-to-noise ratio (SNR), based on the CPU power available
(Brady and Creighton, 1999; Schutz and Papa, 1999).

The algorithms investigated so far deal only with {\it isolated} sources.
To search for a NS orbiting a binary companion has always 
been considered computationally intractable, as one would need to correct also
for the Doppler phase shift caused by the orbital motion of the GW source around a
companion object (star, back hole, or planet): the maximum time of coherent integration,
before the signal power is spread over more than one frequency bin, is between
$\sim 200$ sec (for a source like the Hulse-Taylor binary pulsar) and $\sim 10^5$ sec
(for a NS with a Jupiter-size planet in a 4-months period orbit).
This would add five more search
parameters. Nonetheless, there are several important reasons for start addressing this 
problem at this time:
(i) we would like to quantify what "intractable" means,
and estimate the computational costs 
as a function of the search parameters; (ii) the only known GW source in the 
high-frequency band is Sco X-1, a NS orbiting a low-mass companion (Bildsten, 1998); 
if our current astrophysical understanding is correct, Sco X-1
would be detectable by GEO600 (in narrow-band configuration) and LIGO, at 
a SNR$\simeq 3$ in two years of full coherent integration;
(iii) the continuous monitoring of all known NS's is planned, and around 50 radio 
pulsars are in a binary system; (iv) we are now starting the design of software
codes to search for CW sources during the first science runs carried out by the
detectors, but their general structure is likely to be used for several years.

The aim of this contribution is to estimate the additional processing power
needed to correct for the
orbital motion of a CW source orbiting a binary companion, with emphasis on
targeting known NS's.

\section{Signal model and data analysis}

In order to disentangle the extra computational costs involved in dealing
with the NS orbital parameters, we will
make the following assumptions: the source location in the sky is exactly known -- so 
that one can perfectly remove the  phase Doppler shift due to the 
detector motion -- and the signal is monochromatic, at frequency $f_0$. 
For a general blind search of NS's possibly in binary orbits, the total computational 
burden would be (roughly) the product of the one quoted for isolated sources, times
the estimate that we present here. 

The gravitational waveform is given by
\begin{equation}
h(t, {\vec \lambda}) = \Re \{{\cal A} \,
e^{-i \left[2\pi f_0 t +\phi_D(t; f_0,{\vec \lambda}) + \Psi \right]}\},
\label{h}
\end{equation}
where ${\cal A}$ and $\Psi$ are assumed (as usual) constant, and 
$\phi_D$ is the Doppler phase modulation induced
by the orbital motion of the source around the companion; 
${\mbox{\boldmath $\lambda$}} = (f_0, {\vec \lambda})$ it the signal parameter vector. We
assume that the orbit is Keplerian, and elliptical in shape. 
The Doppler correction to the phase of the signal due to the orbital motion is
therefore:
\begin{equation}
\phi_D (t; f_0,{\vec \lambda}) =  - {2 \pi f_0 a \sin \epsilon \over c}\,
\left[\cos \psi \cos E(t) + \sin \psi 
\sqrt{1 - e^2} \sin E(t)\right]\,,
\label{phid}
\end{equation}
where $\epsilon$ and $\psi$ are the polar angles describing the
direction to the detector, with respect to an appropriate reference frame
attached to the binary system,
$c$ is the speed of light and $E$ is the eccentric anomaly. It is 
related to the mean angular velocity $\omega \equiv 2\pi/P$ (where
$P$ is the orbital period) and the mean anomaly $M$ by the 
Kepler equation: $E - e \sin E = \omega t + \alpha \equiv M$,
where $\alpha$ is an initial phase, $0 \le \alpha < 2 \pi$. 

The additional parameters on which one must launch a search are therefore
five in the elliptical case, say ${\vec \lambda} = (a_p,\omega, \alpha, e, \psi)$,
where $a_p \equiv a \sin \epsilon$, and three in the circular orbit case,
say ${\vec \lambda} = (a_p,\omega, \alpha)$. As usual,
$f_0$ is not a search parameter which requires a filter mesh.

A rigorous way of estimating the search costs can be worked out by approaching
the data analysis through a geometrical picture 
(Sathyaprakash and Dhurandhar, 1991; Dhurandhar and Sathyaprakash, 1994; 
Balasubramanian et al., 1996; Owen 1996;
Brady et al., 1998): 
the signal is a vector in 
the vector space of data trains and the $n$-parameter family of signals traces out
an $n$-dimensional manifold which is termed as the signal manifold.  
On this manifold 
one introduces a proper distance -- and therefore a metric $\gamma_{ij}$ --
defined as the fractional loss of SNR -- the {\it mismatch} $\mu$ -- caused
by the wrong choice of the filter parameters. The spacing of the grid of 
filters is decided by the fractional loss due to the imperfect match
that can be tolerated. Fixing the mismatch $\mu$, 
fixes the grid spacing of the filters in the parameter space $\cal P$. 
The number of filters $N$ is then just:
\begin{equation}
N = \left[\frac{1}{2}\,\sqrt{\frac{n}{\mu}}\right]^{n}\, V_{\cal P}\,,\quad\quad\quad 
V_{\cal P} = \int_{{\cal P}} \sqrt{\det ||\gamma_{ij}||} d {\vec \lambda}\,,
\label{Nf}
\end{equation}
where $V_{\cal P}$ is the proper volume, and $n=5\, (3)$ for eccentric (circular)
orbits.

\section{Computational Costs}

The general expressions of $V_{\cal P}$ and $N$, with signal model given by Eqs.~(\ref{h}) 
and~(\ref{phid}), are very complex. Nonetheless, in the
limit of long ($T\gg P$) and short ($T\ll P$) observation times, 
with respect to the orbital period, one can obtain analytical closed form expressions 
which are actually quite simple. The expansions in these two regimes agree remarkably
well with the exact expression over most of the P-range, see Figure 1.
We give here some details for the circular orbit case, and just sketch the 
key result for eccentric orbits.

\begin{figure}[t]
\begin{center}
\epsfig{file=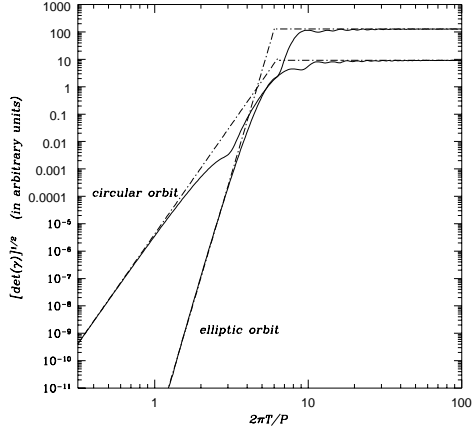,width=0.50\textwidth}
\end{center}
\caption{The proper volume element $\sqrt{det(||\gamma_{ij}||)}$ 
(in arbitrary units) as a function of the time
of observation $T$ in units of the the source orbital period $P$. 
The solid and dashed lines refer to the full numerical expression
and the analytical approximation, respectively (see text).
The plot refers to a binary system with $a_p/(c T) = 3\times 10^{-7}$ and 
$\alpha = 0$.
}
\label{fig1}
\end{figure}

The circular orbit case is important because it provides us several insights into the problem
via a comparatively easier computation; moreover,
several binary radio pulsars have effectively $e=0$ and Sco X-1 is 
essentially in a circular orbit; in addition, when "blind" searches 
will be implemented, it is likely that they will be restricted to 
NS's orbiting a companion in circular orbit, in order to keep the computational 
burden affordable. The total number of filters (for a $3\%$ mismatch) is given by:
\begin{equation}
N \sim
\left\{ \begin{array}{ll}
10^{17}\, \left(\frac{V}{10^{15}}\right)
& \quad\quad (2\pi T/P \gg 1) \nonumber\\
10^{4}\, \left(\frac{V}{100}\right)\,
& \quad\quad (2\pi T/P \ll 1) \nonumber\\
\end{array}
\right.
\,,
\label{N0}
\end{equation}
where the parameter volume one needs to cover is:
\begin{equation}
V \simeq
\left\{ \begin{array}{ll}
1.5\times 10^{15}\,\left(\frac{f_{\rm max}}{1\,{\rm kHz}}\right)^3
\,\left(\frac{a_{\rm max}}{5\times 10^{10}\,{\rm cm}}\right)^3\,
\left(\frac{\omega_{\rm max}}{6.3\times 10^{-4}\,{\rm s}^{-1}}\right)\,
\left(\frac{T}{10^{7}\,{\rm s}}\right)
& (2\pi T/P \gg 1) \nonumber\\
200\,\left(\frac{f_{\rm max}}{1\,{\rm kHz}}\right)^3
\,\left(\frac{a_{\rm max}}{5\times 10^{10}\,{\rm cm}}\right)^3\,
\left(\frac{\omega_{\rm max}}{6.3\times 10^{-4}\,{\rm s}^{-1}}\right)^9\,
\left(\frac{T}{10^{3}\,{\rm s}}\right)^9
& (2\pi T/P \ll 1) \nonumber\\
\end{array}
\right.
\,;
\label{Vmax0}
\end{equation}
here we have ambitiously chosen parameter values that are appropriate for a NS/NS
binary in a few hours orbit. Notice the radically different 
dependence on $\omega T$ in the two regimes, and the $a_p^3$ dependence of the
computational costs. If the parameter values are known in
advance with an error $\pm \delta \lambda^j$, then the number of filters
reduces by a factor
\begin{equation}
\frac{\delta V}{V} \simeq
\left\{ \begin{array}{ll}
2.4\times 10^{-5} \prod_{j=1}^3\,\left[
\left(\delta \lambda^j/\lambda^j\right)/10^{-2}\right] 
&  \quad\quad (2\pi T/P \gg 1) \nonumber\\
2.2\times 10^{-4} \prod_{j=1}^3\,\left[
\left(\delta \lambda^j/\lambda^j\right)/10^{-2}\right] 
&  \quad\quad (2\pi T/P \ll 1) \nonumber\\
\end{array}
\right.
\,.
\label{G0}
\end{equation}
Clearly, any prior information on the value of the source parameters greatly decreases 
the number of templates, and it is easy to check that a single-filter search 
can be performed if :
\begin{equation}
\frac{\delta \lambda^j}{\lambda^j} {\lower.5ex\hbox{$\; \buildrel < \over \sim \;$}} 
\left\{ \begin{array}{ll}
7\times 10^{-7}\,
\left(\frac{V_{\rm max}}{10^{15}}\right)^{-1/3}
&   \quad\quad (2\pi T/P \gg 1) \nonumber\\
8\times 10^{-3}\,
\left(\frac{V_{\rm max}}{100}\right)^{-1/3}
&   \quad\quad (2\pi T/P \ll 1) \nonumber\\
\end{array}
\right.
\,,
\quad\quad(\forall j)\,.
\label{err0}
\end{equation}

In the case of eccentric orbits, the computational burden is of course
much higher, but the results show a behaviour which is very similar to
the one for $e=0$. As in the previous case, the costs increase dramatically
as the observation time covers more than $\simeq 1$ rad of the source orbital
phase. For $P\gg T$ one finds:
\begin{eqnarray}
N^{(e)} & \sim & 10^{26}\, 
\left(\frac{V^{(e)}}{9\times 10^{21}}\right)\,,
\\
V^{(e)} & \propto & f_{\rm max}^5\,a_{\rm max}^5\,\omega_{\rm max}\,T\,
\,F(e)\,;
\end{eqnarray}
$F(e) = e^2 (1 - 3 e^2/8) + O(e^6)$ contains the dependency on the eccentricity,
and $V$ is normalized to the parameter values given in Eq.~(\ref{Vmax0}); 
notice that now, with a 5-dimensional parameter space, the number of filters is 
proportional to $a_p^5$.

\section{Targeting known sources in binary systems}

We can now apply the results of the previous section to some of the known 
NS in binary systems which will be targeted by laser interferometers,
and estimate the processing power involved {\it only} in the coherent correction
of the orbital motion.

Sco X-1 orbits a 
low-mass companion with a period $P = 0.787313(1)$ days, in an orbit
which is essentially circular (here we will assume $e=0$); the position
of the NS on the orbit is known with an error $\simeq 0.1$ rad, and 
the projected semi-major axis is $a_p \simeq 6.3\times 10^{10}$ cm, 
with $\delta a_p/a_p\simeq 5.16\times 10^{-2}$. The uncertainties 
surrounding the frequency at which GW's are emitted suggest to cover
a frequency band up to $\approx 600$ Hz. 
It is easy to verify that for integration times
longer than $\sim 4$ hours, one needs to correct for the source orbital motion.
It is also evident, from the results of the previous section, that the number
of filters changes dramatically when the integration time goes from $\sim 6$ hours to
a day; in fact:
\begin{equation}
N_{\rm Sco} \simeq 
\left\{ \begin{array}{ll}
2.5 \times 10^{6}\,
\left(\frac{f_{\rm max}}{600\,{\rm Hz}}\right)^3\,
\left(\frac{T}{10^{5}\,{\rm sec}}\right)
& \quad\quad (T{\lower.5ex\hbox{$\; \buildrel > \over \sim \;$}} 1\, {\rm day}) \\
14 \left(\frac{f_{\rm max}}{600\,{\rm Hz}}\right)^3\,
\left(\frac{T}{5\,{\rm hour}}\right)^9
& \quad\quad (T{\lower.5ex\hbox{$\; \buildrel < \over \sim \;$}} 6\, {\rm hours}) \\
\end{array}
\right.
\,.
\label{Gc_Ps}
\end{equation}
Notice also that if one allows for possible (small) departures from a perfectly
circular orbit, the number of templates further increases.
 
We analyze now the case of radio pulsars, and assume $T = 10^7$ sec and $\mu = 0.03$.
We have considered
the 44 NS's with a binary companion (the total number of radio pulsars is 706) 
included into the catalogue by Taylor et al. (1993,1995). Seven
binary radio pulsars emit at frequencies below 10 Hz, and are therefore outside
the observational band; for 21 sources the parameter measurements coming from
radio observations are so precise that one can simply use the quoted values
of the parameters and fully correct for the orbital motion; the remaining 
16 radio pulsars require a search over a limited parameter range. For these
sources, all the NS's whose spin-down values yield upper-limits on the GW amplitude 
which are above the
sensitivity curve of all proposed detectors (including LIGO III) would require 
at most 10 templates for $T = 10^7$ sec. Four NS's require a number of templates
in the range $10^2-10^8$ (but are well below the LIGO III sensitivity) and
eight sources, for which we do not have as yet measurements of the spin-down,
would require a very substantial number of orbital filters, more than $\sim 10^{10}$.

\section{Conclusions}

We have presented some preliminary results regarding the additional computational
costs involved in the correction of the Doppler phase shift induced in the
CW signal of a NS orbiting a binary companion; we have given general expressions
to compute the number of filters which are required to carry out the search as
a function of the mismatch, time of integration and parameter space. 
A more thorough description of the issues presented here is currently in 
preparation, and we plan to start soon  the investigation of hierarchical 
algorithms, that would conceivably speed up the search in a considerable way.

\section{References}

\re
1. Balasubramanian R., Sathyaprakash B. S., Dhurandhar S. V. \ 1996, Phys. Rev. D 53, 3033

\re
2. Bildsten L. \ 1998, ApJ 501, L89

\re
3. Brady P. R., Creighton T., Cutler C., Schutz B. F. \ 1998, Phys. Rev. D 57, 2101

\re
4. Brady P. R., Creighton T. \ 1998, gr-qc/9812014

\re
5. Dhurandhar S. V., Sathyaprakash B. F. \ 1994, Phys. Rev. D 49 , 1707

\re
6.\ Owen B. J. \ 1996,  Phys. Rev. D 53, 6749

\re
7. Schutz B. F., Papa M.A. \ 1999, pre-print gr-qc/9905018

\re
8. Taylor J. H., Manchester R. N., Lyne A. G. \ 1993, ApJS 88, 529.

\re
9. Taylor J. H., Manchester R. N., Lyne A. G., Camillo F. 1995, 
{\it Catalog of 706 pulsars} available via anonymous ftp at pulsar.princeton.edu;
unpublished work.

\re
10. Sathyaprakash B. F., Dhurandhar S. V.,  \ 1991, Phys. Rev. D 44 , 3819

\end{document}